\documentclass[aps,pra,a4paper, twocolumn, 10pt, superscriptaddress]{revtex4-1}
\usepackage[T1]{fontenc}
\usepackage[utf8]{inputenc}
\usepackage{lmodern}
\usepackage{graphicx} % Allows including images
\usepackage{amsmath}
\usepackage{amsfonts} % para usar identidad=\mathbb{I}\sqrt{}
\usepackage{amssymb}  % assumes amsmath package installed
\usepackage{mathtools}
\usepackage{multirow}

\usepackage{color}   %May be necessary if you want to color links
\usepackage{hyperref}
\hypersetup{
	colorlinks=true, %set true if you want colored links
	linktoc=all,     %set to all if you want both sections and subsections linked
	linkcolor=blue,  %choose some color if you want links to stand out
}

		{\end{pmatrix}\end{medsize}}%

\newcommand\vect[1]{\overrightarrow{\bf{#1}}} 

\newcommand\conm[2]{\left[ {#1}, {#2} \right] }

\newcommand\parametros[0]{$\kappa/\omega=0.75$, $1.00$, $1.50$, in continuous blue, dashed red, and dotted violet lines respectively}

\DeclarePairedDelimiter\abs{\lvert}{\rvert}%
\DeclarePairedDelimiter\ket{\lvert}{\rangle}%
\DeclarePairedDelimiter\bra{\langle}{\rvert}%
\DeclarePairedDelimiter\expec{\langle}{\rangle}%

\newenvironment{lcase}
{\left\lbrace \begin{aligned}}
	{\end{aligned} \right.}

\begin{document}

\title{A trapped ion in an optical cavity: numerical study of an optomechanical
transition in the few-photon regime}

\author{Alan Kahan}
\affiliation{Instituto de F\'{i}sica Enrique Gaviola, CONICET and Universidad Nacional de C\'{o}rdoba,
Ciudad Universitaria, X5016LAE, C\'{o}rdoba, Argentina}

\author{Leonardo Ermann}
\affiliation{Departamento de F\'isica Te\'orica, GIyA, Comisi\'on Nacional de Energ\'ia At\'omica, Buenos Aires, Argentina}
\affiliation{CONICET, Godoy Cruz 2290 (C1425FQB) CABA, Argentina}

\author{Cecilia Cormick}
\affiliation{Instituto de F\'{i}sica Enrique Gaviola, CONICET and Universidad Nacional de C\'{o}rdoba,
Ciudad Universitaria, X5016LAE, C\'{o}rdoba, Argentina}

\date{\today}

\begin{abstract}

We consider an optomechanical system composed by a trapped ion dispersively coupled to a single mode of a pumped optical cavity. We focus in a parameter range for which the semiclassical description predicts two clearly distinct equilibrium configurations in the limits of small and large photon pumping, while a bistable regime is found for intermediate pumping. This semiclassical description, however, is not valid in close proximity of the system transitions or when the mean photon number is low. Here we provide a numerical analysis of the fully quantum state in the few-photon regime, exploring the features of the asymptotic state across the transition and analyzing possible markers of semiclassical bistability. We find an increase in the entropy of the system and of the entanglement in the transition region, but no clear signatures of metastability in the spectrum of the evolution.

\end{abstract}

\maketitle

\section{Introduction}
 
Trapped ions in optical cavities represent a promising example of a hybrid quantum setup, with several potential applications for quantum information such as coupling of processor and flying qubits or creating entangling gates for photons \cite{sterk2012, steiner2013, Walker_PRL_2018}. The use of strong optical potentials for ions has also been demonstrated as a trapping or pinning mechanism \cite{Schneider_NatPhot_2010, linnet2012}, and proposed as a strategy to reduce unwanted heating effects in experiments involving both neutral and charged particles \cite{Cormick_2011, grier2009observation}. The application of optical forces can be useful for tailoring the structure and motional spectrum of an ion crystal and for studying noise-induced transport, among other proposals \cite{Horak_2012, laupretre2019, Cormick_PRA_2016, cetina2013}.

One of the many appealing features of this kind of hybrid setups is the possibility to realize controllable small-scale models of interest for statistical mechanics and condensed matter. In particular, different groups have proposed and demonstrated the use of this platform to explore features of the Frenkel-Kontorova model for friction and crystal dislocation \cite{Frenkel1938, Braun2013}. This line of work builds upon an early experiment observing the anomalous dynamics of a single ion \cite{Katori_1997}, and recent theoretical and experimental advances have lead to the realization of similar models involving chains with a few ions \cite{Garcia-Mata2007, Pruttivarasin2011, Benassi2011, bylinskii2015, fogarty2015, gangloff2015velocity, bylinskii2016observation}.
 
Another interesting aspect of the combination of ion traps and optical cavities is the possibility to apply cavity cooling techniques to the normal modes of the ion crystals. Indeed, by appropriately choosing the system parameters the cavity losses can provide a dissipation mechanism to reduce the kinetic energy of the atoms. This topic has been studied extensively for neutral atoms \cite{Horak_1997, Domokos_2002} and has been proposed as a means to achieve ground-state cooling of long chains of ions as well \cite{Fogarty_2016}.
 
Optomechanical systems in the regime of sufficiently large couplings are known to lead to multistability (see \cite{aspelmeyer2014cavity} for a review). For the particular case of ion crystals in optical cavities, such behaviour has been studied in connection with the linear-zigzag structural transition \cite{cormick2012structural}, and the sliding-pinned phase transition \cite{fogarty2015}. The former case corresponds to a linear ion chain aligned transverse to the cavity axis, whereas the latter is obtained when the chain is along the propagation direction of the light. Both of these models exhibit parameter regimes where two different classical configurations are stable, and thus the semiclassical treatment predicts an asymptotic state which depends on the initial state. 

The semiclassical description is comparatively simple and properly describes the behaviour of the system in many useful cases, but there are several situations in which this treatment is not reliable. These scenarios include the few-photon regime and the case in which the spatial delocalization of the ion is not small enough compared with the relevant scales of the problem. In particular, the semiclassical description is expected to break down in the vicinity of a transition when a motional mode becomes unstable. The characterization of this regime is an open problem, and previous work in nonlinear optical systems has shown that the semiclassical and the fully quantum solutions may differ in the continuous or discontinuous character of the transition \cite{casteels2017, krimer2019}.

The kind of dynamics we consider is an instance of a driven-dissipative model. 
Several studies in the literature show similar systems in which a semiclassical approximation leads to a dynamical 
map with more than one fixed point \cite{krimer2019,angerer2017,zens2019,halati2019}.
In a quantum treatment, multistability is the result of a vanishing gap in the Liouvillian, whereas metastability is found for gaps that are much smaller than every other relevant time scale, as has been studied in detail in the open Rabi model \cite{le_boite2017,hwang2018} and the open Ising model \cite{rose2016}.
The closing of the Liouvillian gap in the appropriate thermodynamic limit is one of the defining features of a dissipative phase transition \cite{ciuti2018}.

\begin{figure}
\includegraphics[width=0.8\columnwidth]{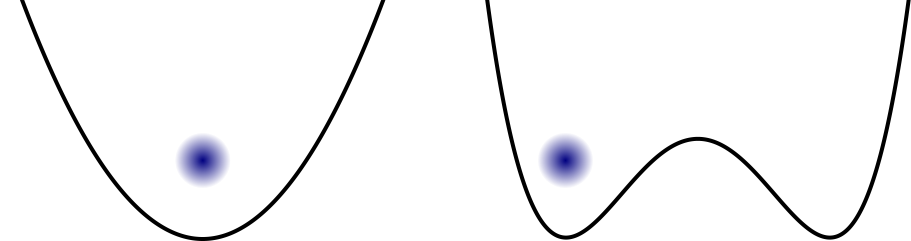}
\caption{\label{fig:scheme} Schematic representation of the different equilibrium configurations in the optomechanical model we consider: a) weak pumping regime, the cavity has a low mean number of photons and the equilibrium for the ion position is located at the center of the ion trap; b) strong pumping regime, the cavity has a large mean number of photons and the equilibrium position of the ion is determined mostly by the optical potential, exhibiting spontaneous symmetry breaking in the semiclassical treatment. If back-action in the system is relevant, the semiclassical treatment may predict stability of both kinds of solutions in the intermediate pumping regime.}
\end{figure}

As a first approach to the full quantum treatment of our system, in this work we study the case of a single motional mode coupled to the cavity field. This mode may correspond to a single trapped ion, or it may refer to the effective description of a problem in terms of a mode which is particularly relevant for the dynamics. The configuration is such that the equilibrium position results from a competition of the trap potential and the optical forces, as depicted in Fig. \ref{fig:scheme}. We study the system in the regime in which the semiclassical description predicts bistability, but in the limit 
of small mean photon numbers, more precisely of about ten photons, and across the whole transition, so that the semiclassical description is not expected to be valid. We provide a characterization of the states of field and ion, as well as joint properties such as entanglement and entropy, across the transition region.

Our results display some features of the transition predicted by the semiclassical treatment in \cite{cormick2012structural}, but no indications of metastability. We confirm the expectation that the transition becomes sharper when the mean number of photons in the regime of classical bistability is increased. We also observe local maxima of logarithmic negativity, mutual information and von Neumann entropy for the asymptotic state of the system in the transition region. We do not detect, however, a decrease in the spectral gap, so that at all times the asymptotic state is clearly separated from the rest. In general, we confirm that the semiclassical description of the state is not appropriate in this regime, but it does capture the right location of the transition region.

This paper is organized as follows: in Section \ref{sec:model} we present the theoretical model for our system. In Section \ref{sec:semiclassical} we briefly review the semiclassical treatment and predictions for equilibrium and bistability. Section \ref{sec:results} contains the results for the quantum regime in the few-photon limit, including studies of the cavity field, the state of the ion and the potential indicators of metastability. In Section \ref{sec:conclusions} we summarize our results and state conclusions and open questions. Finally, several technical issues are explained in the appendices.

%---------------------------------------------------------------------------------
\section{Trapped ion coupled to an optical cavity.} \label{sec:model}

In this section we describe the evolution of a trapped ion coupled to a single mode of an optical cavity pumped by a laser, in the dispersive regime. 
In the reference frame rotating at the laser frequency, the unitary evolution of this system is given by
\begin{equation}
  H =H_{\rm cav}+H_{\rm ion}+H_{\rm int}  
\end{equation}

Here,
\begin{equation}
  H_{\rm cav}=-\hbar \Delta_c a^\dagger a + i \hbar  \left(a^\dagger \eta -a  \eta^* \right)
\end{equation}

\noindent is the intracavity field Hamiltonian. The operators $a^\dagger$ and $a$ are the creation and annihilation operators of the cavity field,
respectively, and $\Delta_c=\omega_l - \omega_c$ is the detuning between the laser pump frequency and the cavity mode, while $\eta$ is proportional 
to the amplitude of the 
laser field. Since the phase of the laser field is arbitrary, we set it to zero for simplicity, hence $\eta \in \mathbb{R}$.

The intracavity field is coupled with modes of the external field. The thermal occupation of the outer modes will be neglected and thus we will assume that this coupling, apart from the pumping, can only cause photon losses. The photon decay, at a rate $2\kappa$, will be described with the standard master equation under the Markov approximation:
\begin{equation}
 \mathcal{L}_\kappa \rho = \kappa (2 a \rho a^\dagger - \{a^\dagger a, \rho\} )
\end{equation}
with the curly bracket denoting an anticommutator.

The Hamiltonian describing the motional degrees of freedom of the ion in absence of the cavity is:
\begin{equation}
  H_{\rm ion} = \frac{p^2}{2m} + V_{\rm ion}(x)
\end{equation}

\noindent Here, we assume that the ion is strongly confined in two directions and thus only the motion along one axis with coordinate $x$ is relevant. % We take this axis to correspond to a direction orthogonal to the propagation direction of the cavity, so that the ion experiences the optical potential associated with the profile of the cavity field.
For our purposes, the potential for the ion includes only the trap potential, but in the more general case of a system 
composed of more than one ion the Coulomb potential should be included. We resort to the usual harmonic approximation:
\begin{equation}
  V_{\rm ion}(x)=\frac{m\omega^2 x^2}{2} \,.
\end{equation}

The interaction Hamiltonian between the intracavity field and the ion degrees of freedom is given in the dispersive regime by: 
\begin{equation}\label{int}
  H_{\text{int}}= \hbar \frac{\Omega^2(x)}{\Delta_0} a^\dagger a
\end{equation}
This approximation is valid as long as 
the laser frequency is far enough from resonance with the electronic transition at frequency $\omega_e$, so that the detuning $\Delta_0=\omega_l-\omega_e$ determines a time scale which is much faster than any other in the system. The frequency ${\Omega}(x)$ in the previous equation is the
coupling strength between the cavity field and the electronic degrees
of freedom of the ion; since this coupling is assumed to be spatially
varying, it also couples the cavity field with the atomic position.
We do not consider spontaneous emission in our description, which is a valid approximation 
as long as $\abs{\Delta_0}$ is sufficiently large \cite{grimm2000optical}. We also note that we do not include additional noise or friction on the ion; it is the interaction with the cavity field that indirectly provides dissipation for the motion as well.

The effect of the interaction Hamiltonian (\ref{int}) can be understood by first considering its effect on ion and cavity separately: if the number of photons $N$ is assumed to be externally fixed, then the coupling with the atom provides an optical potential of the form $\hbar N \Omega^2(x) / \Delta_0$. For $\Delta_0>0$, i.e. a blue-detuned laser, the optical forces push the ion to minima of the optical intensity; if $\Delta_0$ is chosen negative, the optical forces push the ions to locations where the field intensity is maximum. On the other hand, if the ion position $x$ is taken as fixed, it provides a position-dependent shift of the cavity frequency. This can be made more explicit rewriting the total Hamiltonian as:
\begin{equation}\label{Hsys}
  H = - \hbar \Delta_{\text{eff}}(x) a^\dagger a + i\hbar \eta \left( a^\dagger -a \right) + H_{\rm ion}(x)
\end{equation}
\noindent where the operator $\Delta_{\rm eff}$ is defined by 
\begin{equation}
  \Delta_{\text{eff}} (x) = \Delta_c - \frac{\Omega^2(x)}{\Delta_0}\,.
\end{equation}
In the complete scenario, since the effective cavity detuning is relevant to determine the number of photons in the cavity, and in turn the number of photons in the cavity determines the optical potential depth, both degrees of freedom interact in a way that gives rise to ``back-action'' of the ion on the optical forces it feels. 

For strong pumping, one expects the equilibrium positions of the ion to be essentially determined by the optical potential, whereas for weak pumping the trap potential dominates. If the optical potential and the trap potential compete, for intermediate values of the pumping the semiclassical treatment of the problem may find multistability, i.e. a coexistence of different solutions for the ion position which are associated with different values of the mean photon number (this will be analyzed in more detail in the next Section).

Following the lines of previous works, we focus on the case of a Hamiltonian preserving spatial parity, corresponding to a situation where the optical field is aligned with the trap \cite{cormick2012structural, cormick2013}. We note that studies of the sliding-pinned transition for a finite chain have resorted to an aligned setup in order to recover a sharp behaviour at the transition \cite{Benassi2011, fogarty2015}. Under this assumption, the semiclassical description predicts a steady state which is either even, when the equilibrium position is at the trap center, or with spontaneous symmetry breaking when the ion is localized at one side. 

For definiteness, we take $\Delta_0>0$ and we consider an intensity profile as given by:
\begin{eqnarray}
&&\Omega^2(x) = \Omega_0^2 \, f(x) \,,\\ 
&&f(x) = \left[ ( x/x_{\rm eq})^2-  1 \right]^2 \,. 
\end{eqnarray}
We make this choice since the quarctic potential provides a paradigmatic scenario for the transition from single well to double well. It also makes numerical calculations simpler and allows us to derive some analytical results for the classical equilibrium positions, as shown in the next section and in Appendix \ref{sec:classical_stability}. In our case, the classical equilibrium positions are located at $x=0$ for weak pumping, whereas in the limit of infinite mean photon number they lie at $x=x_\text{eq}$. We also fix our choice in such a way that the optical depth per photon associated with the barrier between optical wells is characterized by the parameter $U_0=\Omega_0^2/\Delta_0$. 

Of course, this intensity profile does not describe an experimentally realistic shape, but only the behaviour in the central trap region is relevant for our purposes. Experimentally, a transition similar to the one we study could be observed taking $x$ to be an axis orthogonal to the propagation direction of the cavity and using a tightly focused blue-detuned Gaussian field or a Hermite-Gaussian mode. Alternatively, one could consider $x$ to coincide with the propagation direction of the field and restrict to the parameter regime where only two wells of the optical sinusoidal potential are accessible to the ion. We prefer not to refer to any specific implementation since we expect the qualitative behaviour of our system to be independent of the details of the potential as long as it displays an optomechanical single-to-double well transition. Furthermore, one motivation for our study are transitions involving more than one ion such as the sliding-pinned \cite{cormick2012structural} or the linear-zigzag \cite{fogarty2015}. In these contexts, one can often find one motional mode which ``drives'' the transition, and it is the effective potential for this mode that is most relevant in the description.

A particularly relevant parameter of our model is the dispersive cooperativity $C=U_0/\kappa$, which quantifies the effect of the ion on the cavity field, and thus the back-action of the ion location on the optical potential it feels. The regime of small dispersive cooperativities leads to the usual optical potentials in dipole traps \cite{grimm2000optical}. On the contrary, large values of $C$ correspond to highly deformable potentials which can display multistability and, when several atoms are considered, significant cavity-mediated interactions (see for example \cite{Mottl_2012, Kulkarni_2013, fogarty2015}). In systems containing several atoms, the effective dispersive cooperativity scales with the number of particles.

\section{Semiclassical approximation and classical equilibrium configurations} \label{sec:semiclassical}

The equations of motion for this system can be written in Heisenberg-Langevin form:
\begin{equation}\label{ec_heisenberg_lang}
\begin{lcase}
  \dot a &= i \Delta_{\rm eff}(x) \, a  + \eta -\kappa a + \sqrt{2\kappa}\, a_{\rm in}(t) \\
  \dot x &= \frac{p}{m} \\
  \dot p &= \hbar a^\dagger a \frac{\partial \Delta_{\rm eff}(x)}{\partial x} - \frac{\partial V_{\rm ion}(x)}{\partial x}
\end{lcase}
\end{equation}
\noindent where the operators $a_{\rm in}(t)$ 
can be interpreted as a stochastic external field \cite{gardiner1985}.

Within the semiclassical treatment the system dynamics are considered to be given by small fluctuations around the system's classical equilibrium positions:
\begin{equation}\label{semiclasico}
\begin{lcase}
  a &= \overline{a} + \delta a \\
  x &= \overline{x} + \delta x \\
  p &= \overline{p} + \delta p
\end{lcase}
\end{equation}
\noindent where $\overline{a} \equiv \expec{a}$, $\overline{x} \equiv \expec{x}$, and $\overline{p} \equiv \expec{p}$, 
so that the mean values of the fluctuation operators are zero.
In order to find the classical equilibrium values the input noise fields $a_{\rm in}(t)$ are set to zero, while looking for solutions satisfying $\dot a = \dot x = \dot p = 0$.
This leads to:
\begin{equation}\label{aeq}
  \overline a = \frac{\eta}{\kappa -i \Delta_{\rm eff}(\overline x)}
\end{equation}
\begin{equation}\label{peq}
  \overline p = 0
\end{equation}
\noindent whereas the possible equilibrium positions are determined by the condition that
\begin{equation}\label{xeq}
  \frac{\partial}{\partial \overline x} \left[ V_{\rm eff}(\overline x) + V_{\rm ion} (\overline x) \right] = 0,
\end{equation}
\noindent with $V_{\rm eff}$ the effective potential \cite{Fischer_2001}
\begin{equation}\label{pot_efectivo}
  V_{\rm eff}(\overline x)= \hbar \frac{\abs{\eta^2}}{\kappa}\arctan{\left(-\frac{\Delta_{\rm eff}(\overline x)}{\kappa}\right)} \,.
\end{equation}
This somewhat strange form of the effective potential results from taking into account in a single expression the dependence on $x$ that comes from $\Omega(x)$ in the coupling Hamiltonian (\ref{int}) and also the fact that the equilibrium mean photon number depends on the position $x$ through Eq.~(\ref{aeq}). More details can be found in \cite{Fischer_2001}.

The classical equation (\ref{xeq}) for the equilibrium positions only imposes a zero derivative of the total effective potential, but stability considerations for the quantum fluctuations about equilibrium show that the semiclassical equilibrium positions must correspond, as usual, to minima of the total effective potential. One should keep in mind that the effective potential is useful for finding equilibrium positions but should not be interpreted as a classical potential energy.

From Eq.~(\ref{aeq}) we see that the lowest-order semiclassical expression for the mean photon number is given by 
\begin{equation}\label{photon_number_semiclasico}
  \langle a^\dagger a\rangle \simeq \left| \overline a \right|^2 = \frac{\eta^2}{\kappa^2 + \Delta_{\rm eff}^2(\overline x)} \,.
\end{equation}
Using the above expression to approximate the mean photon number implies neglecting fluctuations, which are expected to represent an important contribution to the mean photon number when the value predicted by Eq.~\eqref{photon_number_semiclasico} is small.

Given that the spatial profile of the chosen field intensity is even, we have that 
$\overline x = 0$ is always a solution of Eq.~(\ref{xeq}). Considering the form of the effective optical potential (\ref{pot_efectivo}), it is clear that if there are other equilibrium solutions, they must satisfy $\overline{x}^2<x_{\rm eq}^2$; the equality can be satisfied only for vanishing trap potential or in the limit of infinite pumping strength. The equation for the solutions at the sides, which exist only for strong enough pumping, is given in Appendix \ref{sec:classical_stability}.

From the second derivative of the total effective potential $V=V_{\rm eff}+ V_{\rm ion}$ one can determine parameter conditions for which the semiclassical solutions $\overline x$ transition from stable to unstable equilibrium configurations (see Appendix \ref{sec:classical_stability} for details). This description predicts a stable
equilibrium position at $\overline x=0$ for weak laser pumping, whereas for large pumping strength the stable equilibrium positions approach $\pm x_{\rm eq}$.

\begin{figure}[t]
\centering
    \includegraphics[width=0.75\columnwidth]{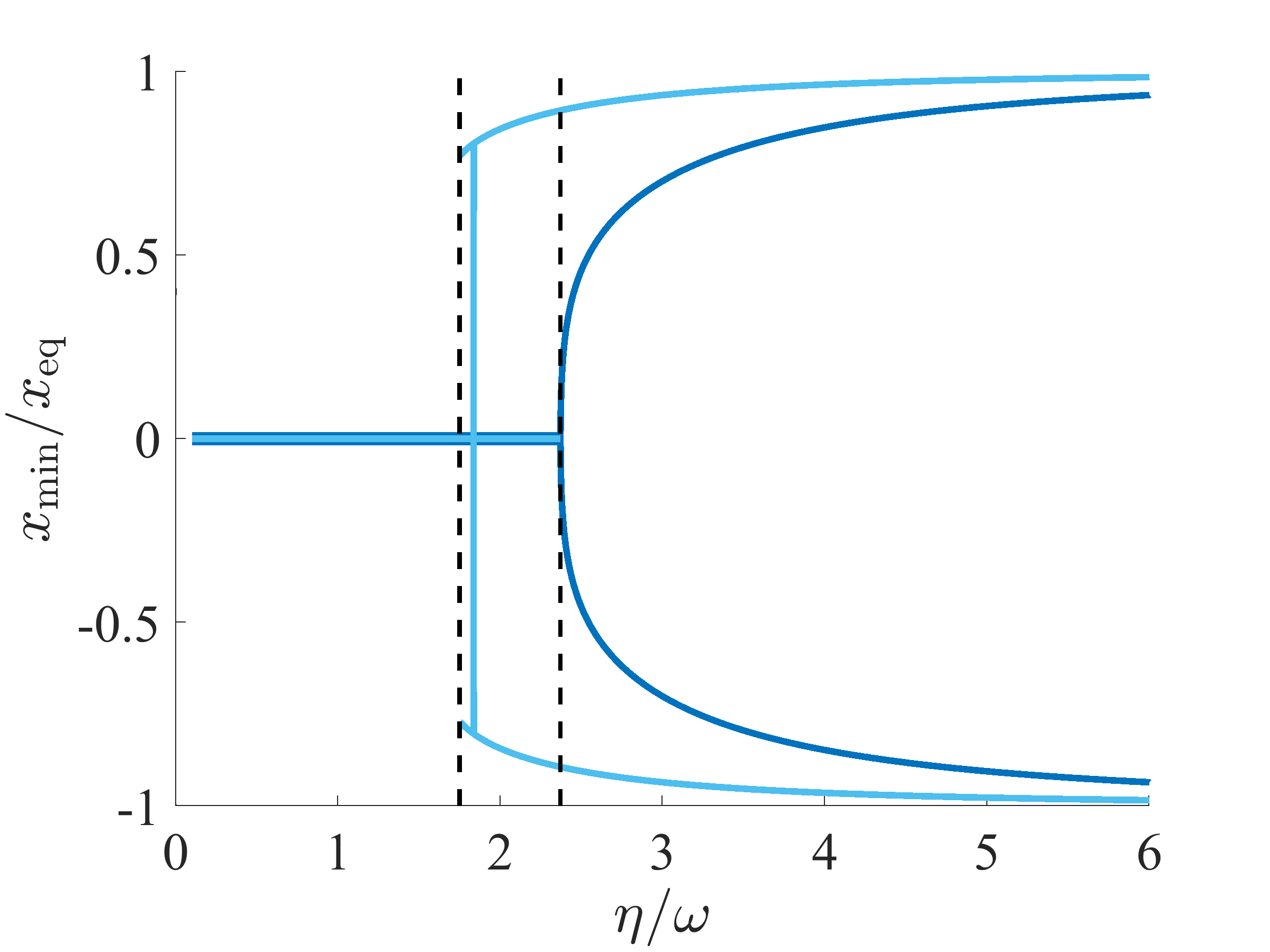}
  \caption{Minima of the total effective potential $V=V_{\rm eff}+ V_{\rm ion}$ determining the equilibrium position in the semiclassical description. 
  The parameters used are $\kappa=\omega$, $x_{\rm eq}=3 \sqrt{\hbar/(m\omega)}$, $c=1$, $C=0.5$ (blue/dark grey), $C=2$ (light blue/grey), corresponding to continuous and discontinuous transitions respectively. For the latter case, the vertical solid line shows the change in global minimum and the
  vertical dashed lines show the transition values computed from Eq. (\ref{segunda_estrellita}) and Eq. (\ref{primera_estrellita}). 
  }
  \label{fig:minimos_pot_eff_transicion}
\end{figure}

The transition between both classes of solutions can be continuous or discontinuous depending on the cooperativity $C$ and a dimensionless parameter $c$ associated with the detuning:
\begin{equation}
 c=-\frac{\Delta_c}{U_0} +1 \,.
\end{equation}
The change from a continuous to a discontinuous transition can be identified by imposing that the instability of the equilibrium points at the sides occurs at a position that approaches zero. This leads to a critical value for $C$ which depends on the detuning:
\begin{equation}
 C_{\rm crit} = \frac{1}{\sqrt{c\,(4-c)}} \,.
\end{equation}
If $C$ is above $C_\text{crit}$, then the transition is discontinuous; otherwise, it is continuous. 
For a detuning such that $c\not\in (0,4)$, the semiclassical description predicts a continuous transition for all values of the cooperativity.

As an illustration, in Fig. \ref{fig:minimos_pot_eff_transicion} we show the minima 
of $V$ as a function of $\eta$, for a cooperativity 
below (dark blue) and above (light blue) the critical value. In the first case a continuous transition can be observed. In contrast, the second case displays an abrupt transition where two local minima
appear close to $x_{\rm eq}$, while the value $\overline x=0$ remains a stable solution, and as $\eta$ increases even further, the local minimum at the origin becomes unstable and the 
only stable solutions are the ones around $x_{\rm eq}$. 

The semiclassical description in \cite{cormick2012structural, fogarty2015} also includes a linear treatment of the fluctuations, truncating the Eqs.~(\ref{ec_heisenberg_lang}) to first order in the displacements from the mean values. From this linear set of equations, if the parameter regime corresponds to a stable configuration, one can find an asymptotic Gaussian state \cite{ferraro2005} for the system. Within this semiclassical description, it is the coupling of motional with field fluctuations that leads, for appropriate parameters, to cavity cooling of the ion. This cooling mechanism is most efficient for large coupling in resonance conditions, facilitating the transfer of vibrational energy to cavity fluctuations \cite{Fogarty_2016}. However, there are also parameter values for which the pump induces heating of the motion and thus the system becomes unstable \cite{cormick2013}. 

In the equations for the fluctuations of cavity and motion taken to first order, the coupling between the two degrees of freedom is proportional to the derivative of the field intensity at the ion position. This means that, for the model we analyze, this optomechanical coupling is zero in the whole parameter region for which the equilibrium position is at the origin. This fact has two important consequences: first, in this regime a direct dissipation mechanism on the ion must be included in order to obtain an asymptotic state. Second, this asymptotic state displays no correlations, quantum or classical, between cavity and ion. These features are due to the linear approximation and not properties of the actual quantum system.

Finally, we note that the semiclassical analysis suggests that the main features of the transition depend on very few dimensionless parameters. This need not be the case in the truly quantum regime. For instance, if the minima of the optical potential, $\pm x_{\rm eq}$, are located within the spatial region occupied by the ground state of the trap, then one cannot expect an abrupt transition between the two kinds of solutions, because the quantum treatment shows that they overlap. The semiclassical treatment is expected to become more accurate in the limit with a large number of photons and with $x_{\rm eq} \gg x_\omega$, where $x_\omega=\sqrt{\hbar/(m\omega)}$ is the length scale determined by the trap ground state. Nevertheless, even under these assumptions, the semiclassical description breaks down close to the instability points due to the large position fluctuations.

\section{Quantum steady state - numerical results} \label{sec:results}

In this section we undertake a fully quantum treatment of the system.
In particular, we study the steady state of the system in the regime in which the semiclassical description predicts bistability, but in the limit of small mean photon numbers, where
the semiclassical approximation is not expected to remain valid. For different parameter choices within this limit, we provide a characterization of the transition and analyze possible indications of the semiclassical bistability through the entropy of the state, the entanglement between field and motion, and the spectral gap. 
The results are obtained by numerical diagonalization of the evolution superoperator, truncating the bases for motional and cavity states; for more details see Appendix \ref{sec:methods}.

\begin{figure}[t]
\centering
    \includegraphics[width=0.65\columnwidth]{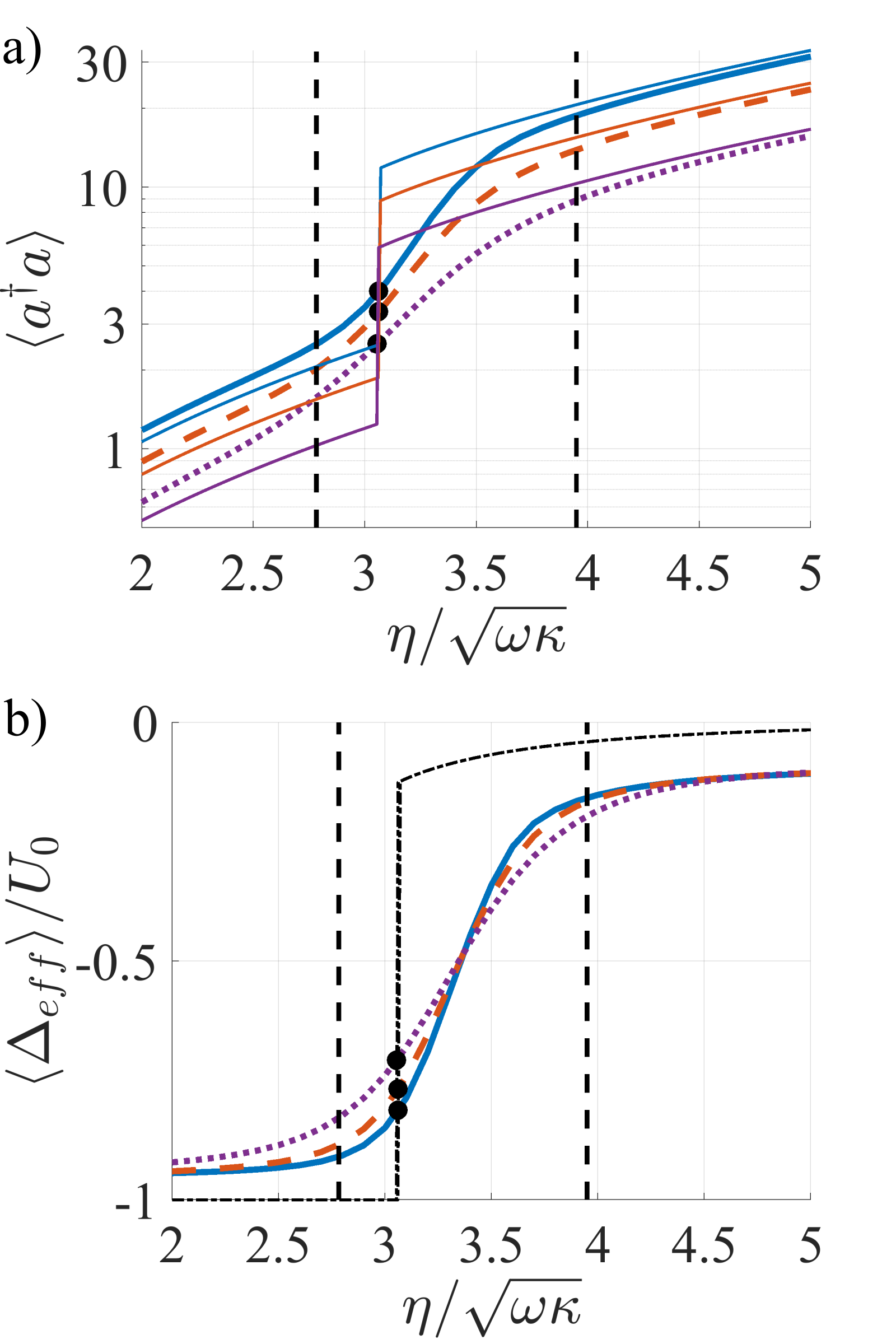}
  \caption{a) Mean photon number (in semilog scale) and b) mean value of the detuning, as a function of the pump strength.  
    We show the mean values obtained through diagonalization of the Liouvillian for the parameter values: $C=2$, $x_{\rm eq}/x_\omega=5$, \parametros. The region for which semiclassical bistable behaviour occurs is shown between dashed vertical black lines, and circles indicate the change in the global classical minimum. In (a) the continous curves with a jump at the global minimum change indicate the values obtained through the semiclassical approximation to lowest order according to \eqref{photon_number_semiclasico}, without taking into account fluctuations. In (b) the dash-dotted line indicates the semiclassical value of the detuning, $\Delta_{\rm eff}(\bar x)$; this curve is the same for the three cases plotted.   
    }
  \label{fig:test_22_means_2}
\end{figure}

\begin{figure*}[htpb!]
\centering
  \includegraphics[width=1\textwidth]{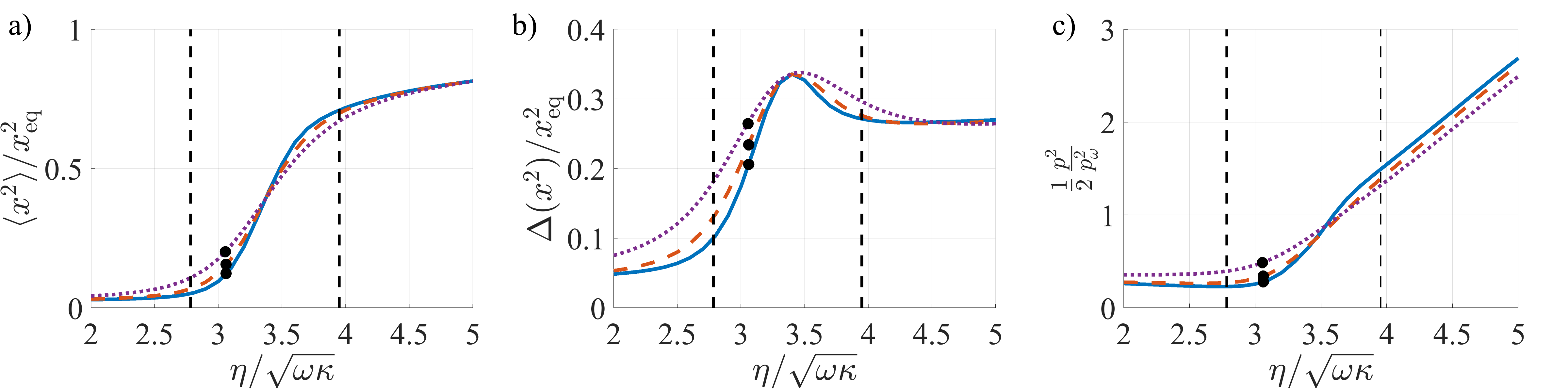}
  \caption{ 
  a) Mean value of the operator $x^2$ associated with the ion position, in units of $x_\text{eq}^2$. We note that the mean value of $x$ is always zero for symmetry reasons.
  b) Dispersion of $x^2$, $\Delta(x^2)=\sqrt{\langle x^4\rangle - \langle x^2\rangle^2}$, in units of $x_\text{eq}^2$. c) Mean value of the kinetic energy of the ion, in units of $p_\omega^2/m=\hbar \omega$. The curves correspond to the parameter values: $C=2$, $x_{\rm eq}/x_\omega=5$, \parametros. Notice that as the cooperativity is kept fixed, different values of $\kappa$ correspond also to different values of $U_0$.
  }
  \label{fig:test_22_means_x}
\end{figure*}

It follows from the symmetry of the Hamiltonian that, unless direct dissipation or noise on the motion is included, the system will have at least two steady states corresponding to subspaces with even and odd parity in the ion's coordinate space \cite{halati2021}. 
The spontaneous symmetry breaking predicted by the semiclassical model can only be observed in the 
quantum description if the tunneling between potential wells is zero, corresponding to the limit of infinitely many photons. 
To simplify the description, and since symmetry breaking is not the feature we want to analyze, we only consider the subspace of even motional states. 

For definiteness we choose $\Delta_c=0$; this choice guarantees $\Delta_{\rm eff}<0$, which was identified as necessary to avoid cavity heating \cite{cormick2013}. Fixing $\Delta_c=0$ also leads to some simple analytical results for the classical configurations, as shown in Appendix \ref{sec:classical_stability}.
We study the steady state for fixed values of cooperativity and of $x_\omega/x_{\rm eq}$. In particular, in the examples that follow we take $C=2$, $x_{\rm eq}=5x_\omega$. These values were chosen so that they display features of the transition of interest and at the same time are numerically tractable and close to the experimentally accessible regime with present technology. Although we do not aim to describe any particular realization, our parameter choices could for instance correspond to calcium ions in the transition at 854 nm, taking $\kappa, U_0$ and $\omega$ of the order of $2\pi \times 100$ kHz \cite{meraner2019}. 

We have explored other sets of parameters, confirming the expectation that smaller photon numbers and smaller values of $x_\text{eq}$ lead to smoother transitions. While the main features of our results are representative of the phenomena we want to study, some aspects of the results shown are not universal: for instance, there are regimes for which the cavity cooling effect is too small to lead to a motional state which is localized with respect to the scale set by $x_\text{eq}$, and also the precise positions of the peaks in quantities such as entropy or negativity vary from one parameter choice to another.

\subsection{Transition in the equilibrium configuration - mean values for cavity and ion}

In Fig.~\ref{fig:test_22_means_2} we show the behaviour of expectation values associated with the cavity field as the pumping is increased. Since the semiclassical bistability region depends on the pump strength through the combination $\eta^2/(\omega\kappa)$, the curves are shown as functions of $\eta/\sqrt{\omega\kappa}$. In order to keep $C$ constant, each curve with constant $\kappa$ has its corresponding value of $U_0$. In every figure in this section, the dashed lines divide the three regions that have a different number of local minima in the total effective potential.
In the leftmost region, $V$ has only one minimum at $x=0$; in the rightmost region only the minima at the sides are stable; the middle region corresponds to classical bistability. 
The change from a global minimum in $x=0$ to one at the sides is shown with small black circles.

Figure \ref{fig:test_22_means_2}-a) displays the mean photon number at the steady state as a function of the pumping and for different values of the cooperativity and of $\kappa,~U_0$. In accordance with the semiclassical description, the mean photon number for fixed $\eta/\sqrt{\kappa\omega}$ grows as $\kappa$ decreases, and outside the transition region the mean photon number grows quadratically with $\eta$. 
The mean photon number obtained through the semiclassical treatment (\ref{photon_number_semiclasico}), 
without taking quantum fluctuations into account, is shown in the same figure with narrow lines that jump at the change of global minimum. One can see that the value predicted from the semiclassical approximation has a different shift at each side of the transition. 
This shift is related with the fact that the semiclassical approximation as described in Sec. \ref{sec:semiclassical} neglects the effect on $\Delta_\text{eff}$ of the spatial spread of the ion. In a fully quantum treatment, this spread has a large impact on the value of the effective detuning and through it on the mean photon number.

The behaviour of $\expec{\Delta_{\rm eff}}/U_0$ as a function of $\eta/\sqrt{\kappa\omega}$ in shown in Figure \ref{fig:test_22_means_2}-b). In dashed-dotted line we display the semiclassical prediction corresponding to $\Delta_{\rm eff}(\bar x)$; there is only one semiclassical curve, since the result does not change when one varies $\kappa$ and $U_0$ keeping $C$ constant (see Appendix \ref{sec:classical_stability} for more details). There is a marked discrepancy between the semiclassical prediction and all the curves obtained from the numerical calculations. This is, as already mentioned, due to the spatial spread of the ion position. 
One can observe in the figure that $\expec{\Delta_{\rm eff}}/U_0$ grows towards $\Delta_c=0$ as $\eta$ increases. This is expected since as $\eta$ is increased, 
the probability density of the ion tends to localize near the potential wells, which in our case correspond to field nodes, reducing the ion-induced cavity detuning. However, the effective detuning reaches a plateau at a value lower than $\Delta_c$, because the ion does not localize completely. One can also confirm from the plot that an increase in the photon number at the transition region gives rise to a more abrupt behaviour.

\begin{figure*}[htpb!]
\centering
    \includegraphics[width=0.85\textwidth]{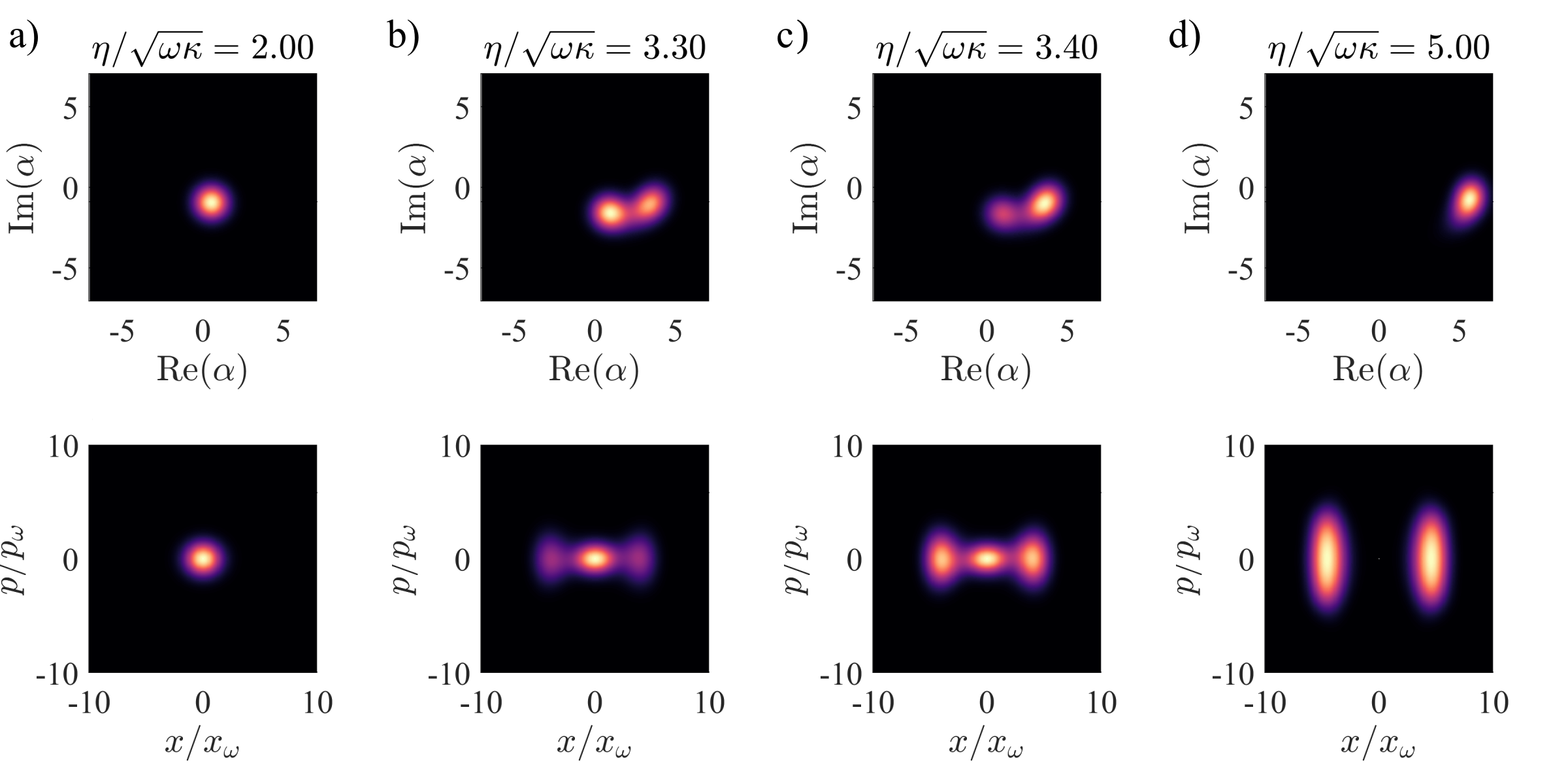}
  \caption{Husimi distribution functions for the reduced states of cavity and ion, for $C=2$, $x_{\rm eq}/x_\omega=5$, $\kappa/\omega=0.75$, and     $\eta/\sqrt{\omega\kappa}=$2.0, 3.3, 3.4, and 5, from left to right. 
  The top row corresponds to the field subsystem, with $\ket{\alpha}$ the coherent states in the Husimi distribution, whereas the bottom row corresponds to the ion.
  }
  \label{fig:test_22_husimis_C2}
\end{figure*}

A deeper understanding of the behaviour of the ion motion can be gained from the several plots in Fig. \ref{fig:test_22_means_x}. The mean value of $x^2$ is displayed in  \ref{fig:test_22_means_x}-a).
Although the curves are qualitatively similar, one can observe that the change in the motional state becomes more abrupt as the values of $\kappa$ and $U_0$ are decreased so that the photon numbers in the transition region become larger. The dispersion of $x^2$, $\Delta(x^2)=\sqrt{\langle x^4\rangle - \langle x^2\rangle^2}$, is shown in Fig.~\ref{fig:test_22_means_x}-b), where one can observe a maximum dispersion around the middle of the transition region. For large pumping, the plots confirm that the ion does not localize completely, and actually the curves become almost flat. This can be understood taking into account that the localization of the ion at the field nodes makes the effective detuning very small, which limits the cooling effect of the cavity. Finally, in Fig.~\ref{fig:test_22_means_x}-c) we show the ion's kinetic energy, in units of $p_\omega^2/m=\hbar \omega$. The corresponding curves are quite flat in the regime with the ion located at the center and display an approximately linear increase with $\eta$ for large pumping. This growth is expected since the motional energy scale associated with the classical optical potential, neglecting back-action and replacing the operator for the number of photons by the mean value, increases like $\hbar\eta$.

\begin{figure*}[htpb!]
\centering
    \includegraphics[width=1\textwidth]{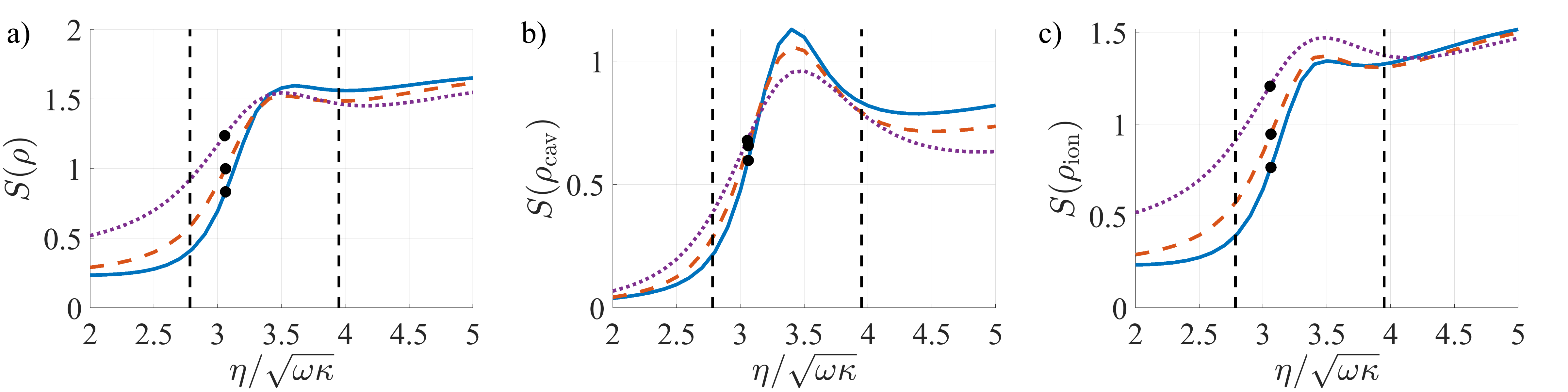}
  \caption{Von Neumann entropy of the composite system and that of the reduced states $\rho_A$ and $\rho_B$ for cavity and ion, respectively. Parameters are the same as in the previous Figures. }
  \label{fig:test_22_purity}
\end{figure*}

In the semiclassical description, the two different equilibrium configurations for the ion directly correlate with two distinct equilibrium values of
the cavity field amplitude. This behaviour can be approximately observed in the Husimi distributions $Q_\rho(\alpha)=\bra{\alpha} \rho \ket{\alpha}$ with $\ket{\alpha}$ a coherent state of the corresponding subsystem \cite{LEE}. We show these distributions in Fig.~\ref{fig:test_22_husimis_C2} for cavity (top) and ion (bottom), with increasing pumping from left to right. In \cite{casteels2017,halati2019}, the authors observe that the Husimi distribution has two distinct lobes in the bistable regime. 
In this sense, we would expect classical bistability to translate into a Husimi distribution for the ion with distinct peaks at $x=\pm x_\text{eq}$ and $x=0$ and with two separated peaks for the associated cavity amplitudes. These conditions are related with the already mentioned requirements that $x_{\rm eq}$ be much larger than $x_\omega$ and the mean photon number much larger than 1. This limit is not reached for our parameter regimes, which is clear from the Husimi functions showing a non-negligible overlap between the peaks that correspond to different semiclassical solutions.

\subsection{Analysis of typical transition markers}

In the following we study the behaviour of several quantities that are expected to display clear signatures of the transition and/or indicate metastability in the quantum description of the system. 
First, we analyze the mixedness of the state, which is expected to have a peak at the transition. For instance, in \cite{ciuti2018} the authors note that the maximum mixedness occurs at the transition between different phases, and in \cite{halati2019}, using the quantum trajectories formalism two different types of trajectories are identified, explaining the asymptotic state as a statistical combination of the two competing equilibrium configurations. In Fig.~\ref{fig:test_22_purity} we show the von Neumann entropy $S(\rho)=-\text{tr}\left(\rho \log \rho \right)$ of the composite system state and that of the reduced states. We observe a clear peak for the cavity, but not in the entropy of the full system or the ion subsystem. Once more this is a consequence of inefficient cavity cooling. 

\begin{figure}[t]
\centering
    \includegraphics[width=0.65\columnwidth]{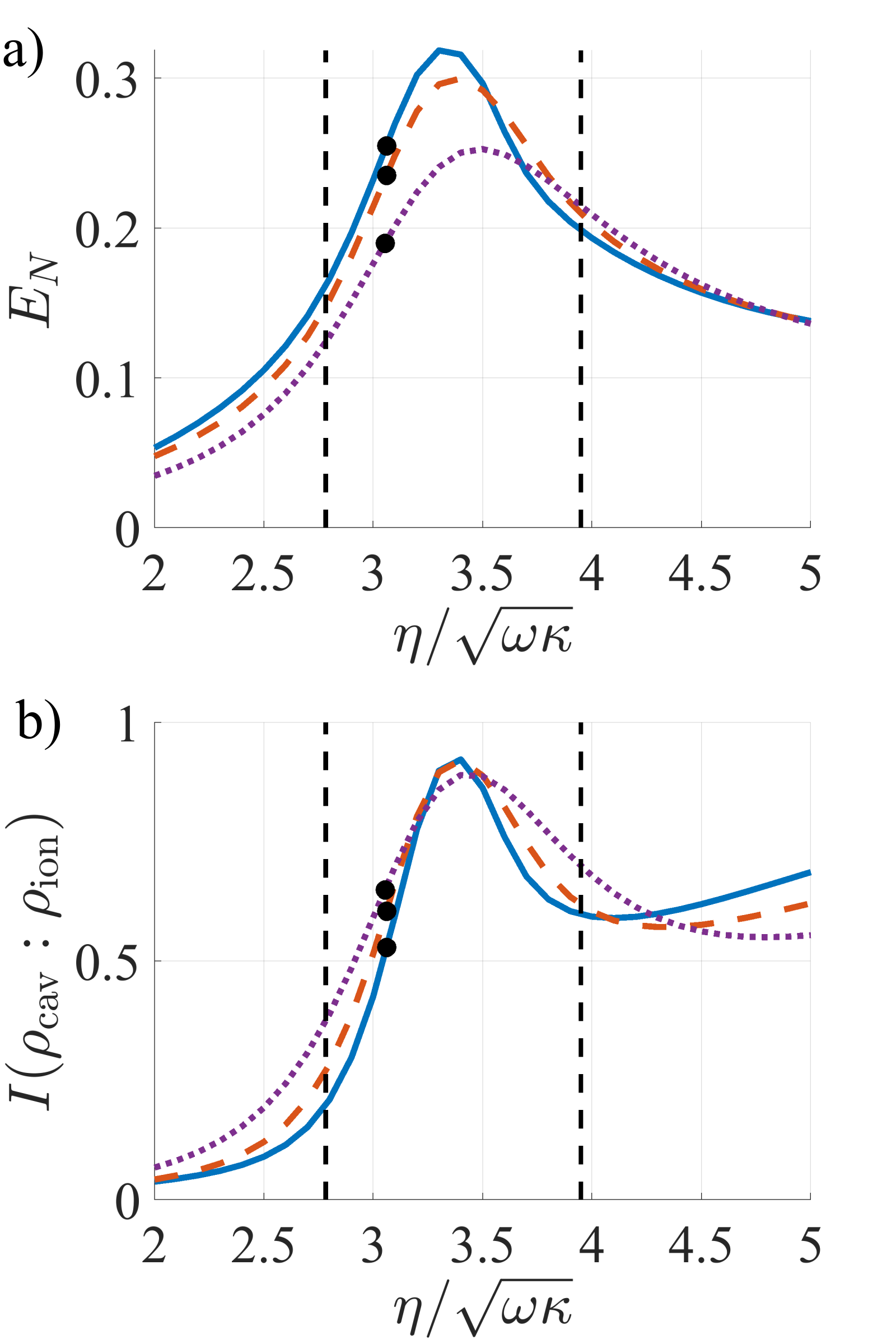}
  \caption{
    a) Logarithmic negativity between motion and cavity degrees of freedom.
    b) Mutual information between motion and cavity degrees of freedom.
    Parameters are the same as in the previous Figures. }
  \label{fig:entanglement}
\end{figure}

Despite the mixedness, there is non-zero entanglement between the ion and the cavity field.  In Fig. \ref{fig:entanglement}-a) we show the behaviour of the logarithmic negativity $E_N$ \cite{werner2002, adesso2007} for the same parameters as in the previous Figures. We note that $E_N$ is calculated from the numerically found steady state $\rho$ as:
\begin{equation}
E_N(\rho) = \log_2 \left(1+2\sum_j |\lambda_j|\right)
\end{equation}
where $\lambda_j$ indicates the negative eigenvalues of the partial transposition of $\rho$. A non-vanishing value of $E_N$ is an indicator of entanglement. For the parameter regimes we study, the negativity reaches a maximum within the transition region. Furthermore, the peak becomes sharper as the mean photon number at the transition increases. In any case, one must be cautious regarding the quantification of entanglement in this kind of system, since different entanglement measures may be more relevant depending on the purpose of the study.

A useful quantifier of the correlations between cavity field and ion motion, which encompasses both quantum-mechanical and classical correlations, is given by the mutual information \cite{nielsen_chuang}. In Fig. \ref{fig:entanglement}-b) we show the behaviour of this quantity, which also displays a peak in the transition region. Interestingly, within the transition region the magnitude of the correlations in terms of the mutual information is roughly the same for all the curves shown, regardless of the difference in mean photon numbers. The large values of the mutual information for strong pumping are related with the shape of the Husimi distribution of the cavity, which exhibits a ``tail'' reminiscent of the alternative equilibrium configuration.

The approach to classical bistability is expected to translate, in a quantum treatment with large numbers of excitations, into metastability, i.e. the appearance of a very slow decay towards the asymptotic state. This time scale, much slower than the other scales in the system, can be observed as a very small gap in the spectrum of the evolution superoperator: the eigenvalue associated with the slowest decay becomes very close to zero. Then, the dynamical relaxation takes place in two steps, namely a fast relaxation towards the metastable manifold followed by a very slow approach to the true asymptotic state \cite{macieszczak2016}. 

\begin{figure}[t]
\centering
    \includegraphics[width=0.65\columnwidth]{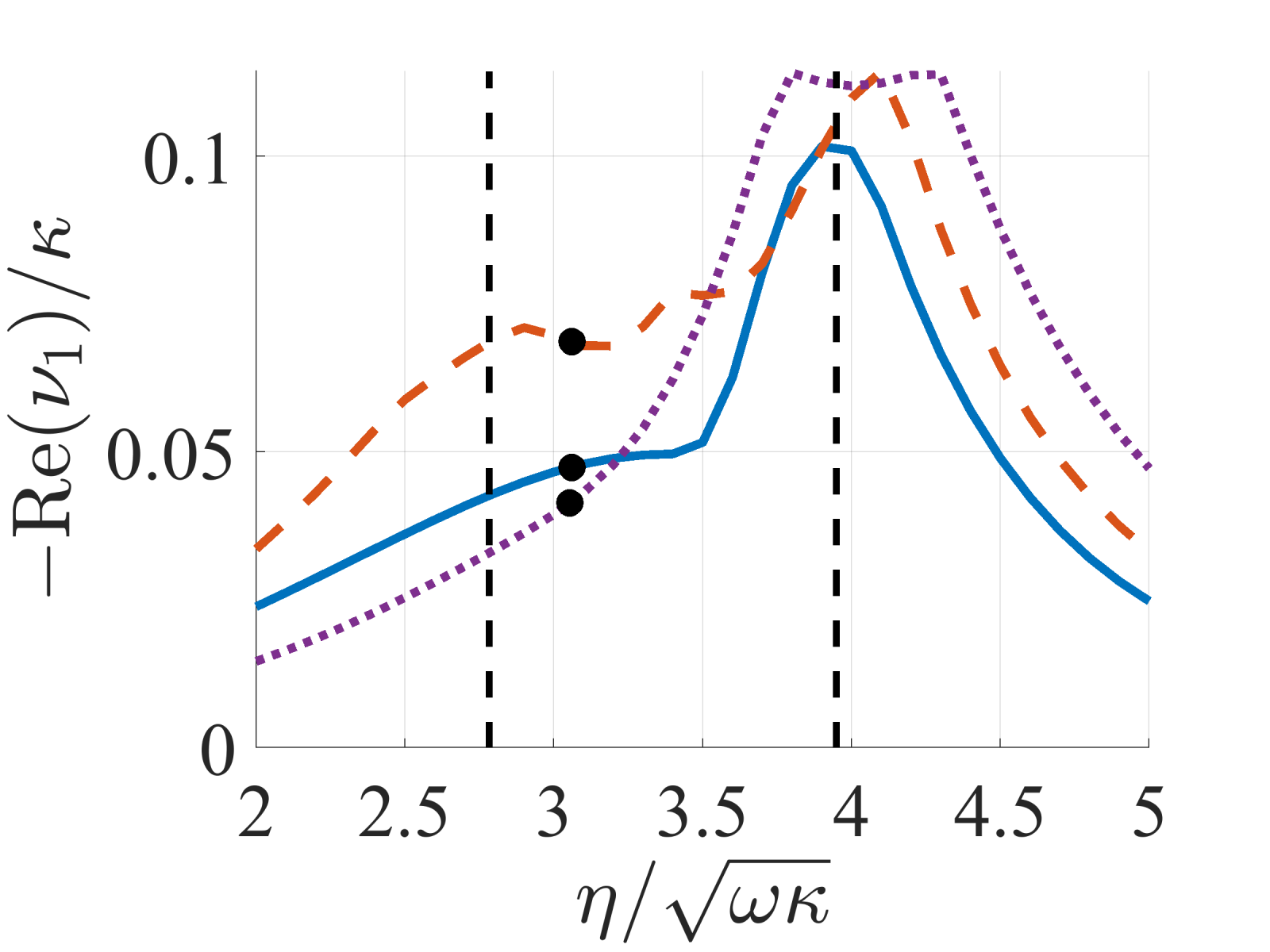}
  \caption{Spectral gap of the Liouvillian in units of $\kappa$. Here, $\nu_1$ is the eigenvalue which is closest to $\nu_0=0$, associated with the steady state. Parameters are the same as in the previous Figures.
    }
  \label{fig:test_22_gap}
\end{figure}

In Figure \ref{fig:test_22_gap} we show the spectral gap of the Liouvillian. We do not observe any significant decrease of the spectral gap in the transition region. However, for the smallest choice of $\kappa, U_0$ there is a sharp change in behaviour around $\eta/\sqrt{\omega\kappa}\simeq 3.5$ which might be a precursor of a gap decrease for higher photon numbers. In general, we rather observe that the gap becomes smaller the further one is from the transition, due to a reduction in the cavity cooling efficiency. The irregular features observed in some of the curves are due to level crossings. We confirm from this plot that larger numbers of photons and/or values of $x_\text{eq}$ are required to approach metastability. Nevertheless, we observe a very slow numerical convergence of the gap values, so that obtaining an accurate estimation of the gap is much more challenging than the calculation of the properties of the steady state. 

\subsection{Gaussian character of the steady state}

\begin{figure*}[t!]
\centering
    \includegraphics[width=0.85\textwidth]{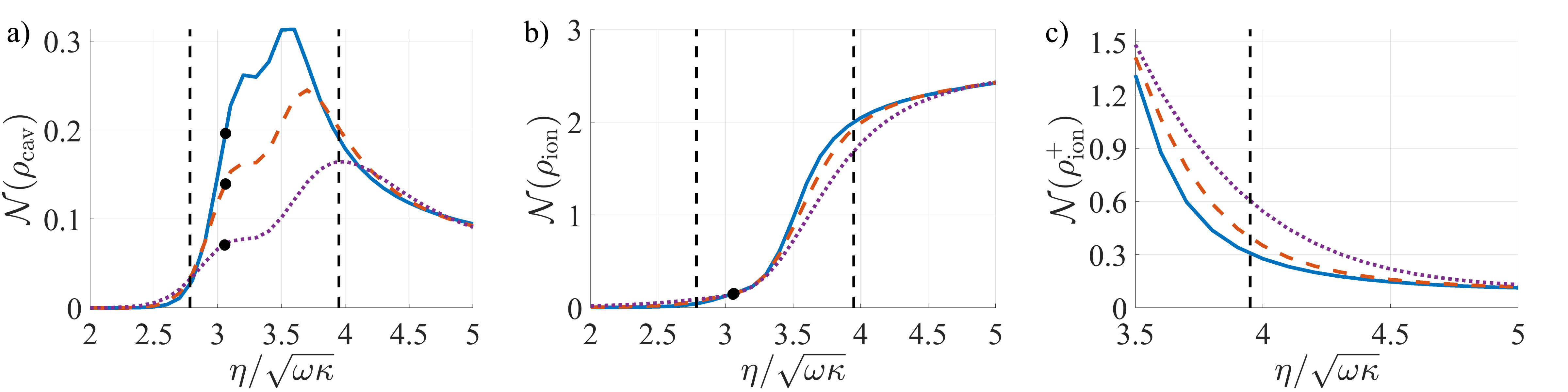}
  \caption{Non-Gaussianity measure $\mathcal N(\rho)$ from Eq.\eqref{eq:non_gaussianity_wehrl}, for a) the cavity state, b) the motional state of the ion, c) the motional state projected on the positive semi-axis (only shown for the parameter region for which the ion is predominantly located at the sides). Parameters are the same as in the previous Figures.
  \label{fig:non-gaussianity-1} }
\end{figure*}

As a final issue, we consider to what extent the asymptotic state can be approximated by a Gaussian state. The semiclassical description can only be approximately correct when the state is close to being Gaussian, although this is a necessary but not sufficient condition. As a quantifier of the non-Gaussian character of the state, we choose to work with the measure defined in \cite{solomon2012}. This choice has several desirable properties, in particular, it focuses on the defining features of non-Gaussianity, namely the higher-order cumulants and its manifestation in the shape of the state in phase-space, providing a quantifier that is invariant under homogeneous scalings in phase space. The measure is based on the Wehrl entropy $H_W$:

\begin{equation}\label{eq:non_gaussianity_wehrl}
  \mathcal N (\rho) = H_W(\rho_G) - H_W(\rho)
\end{equation}

\noindent where $\rho_G$ is the Gaussian state that has the same first and second moments as $\rho$ (for more details see Appendix \ref{sec:gaussian}). For reference, the Fock state $\ket{1}$ is assigned a non-Gaussianity $\mathcal{N} (\ket{1}\bra{1})\simeq 0.12$ \cite{solomon2012}.

In Figs. \ref{fig:non-gaussianity-1}-a) and b) we show this measure for each subsystem as a function of the pumping strength. We resort to the calculation of non-Gaussianity for the subsystems alone since the full system state can only by approximately Gaussian if the subsystem states are. Fig.~\ref{fig:non-gaussianity-1}-a) shows that, as expected, the cavity state is maximally non-Gaussian in the transition region, and is very well approximated by a Gaussian for low pumping. However, this is not the case to the right of the transition; this can be related with the existence of a significant deformation in the corresponding Husimi distribution of Fig.~\ref{fig:test_22_husimis_C2} (a Gaussian state gives an elliptic shape for the Husimi function).  

The non-Gaussianity of the ion state shows a monotonically increasing behaviour with increasing pumping. This can be understood taking into account the spatial symmetry: in the limit of large pumping, our model predicts a probability distribution for the ion with two peaks, one at each side. A true semiclassical state would correspond to one of these two peaks. In order to correct for this effect, and analyze the Gaussianity of just one peak, we consider a state which is projected on the positive (or negative) semi-axis $x$. We achieve this by means of the projection operator such that:
\begin{equation}\label{proyeccion}
  P_+ \ket x = 
    \begin{lcase}
       & \ket x & \text{if} \; x \ge 0 \\
       & 0      & \text{if} \; x  < 0
    \end{lcase}
\end{equation}
followed by the corresponding renormalization of the state. The application of $P_+$ on a state which does not vanish at the origin leads to discontinuities which are unphysical and make the behaviour of momentum ill-defined. However, we work with a modified operator that is truncated in Fock basis. This makes it no longer a proper projection operator, but it gives a qualitatively correct approximation of the projected state avoiding a sharp behaviour at the origin. After this procedure, in the limit of large pumping, the steady state solutions resemble the semiclassical solution with spontaneously broken symmetry. The non-Gaussianity of the resulting state is shown in Fig~\ref{fig:non-gaussianity-1}-c). Using the projected state, the motional state does show a tendency towards a Gaussian character as the pumping strength is increased, indicating that a Gaussian approximation may be applicable far enough from the classically bistable regime.

\section{Conclusions} \label{sec:conclusions}

We have analyzed the asymptotic quantum state of an optomechanical system comprised by the motion of an ion coupled to an optical cavity field, for cooperativities such that the semiclassical treatment predicts bistability, but in the few-photon regime. Our results indicate a smooth transition between the two semiclassical solutions, which is to be expected given the non-negligible overlap between the states associated with the different semiclassical solutions. We observe peaks in the entanglement, mixedness of the cavity state, and mutual information at the steady state in the transition region. However, we do not find clear signatures of the semiclassical bistability in the spectral gap of the evolution superoperator. Instead, the behaviour of the Liouvillian gap is mostly dominated by the efficiency of the cavity cooling of the motion, which is determined by the effective detuning and the ion location.

The semiclassical predictions as in \cite{cormick2012structural} can only be approximately valid when the asymptotic state is close to being Gaussian. According to our results, for the parameters we choose the state can be taken as approximately Gaussian for weak pumping, whereas recovering an approximately Gaussian state in the regime with the ion localized at the sides requires much larger pumping strengths than the ones for which the transition is observed. 
Thus, for the parameter range we explore we would expect the semiclassical description to be approximately valid only in the low-pumping side of the transition. However, contrary to the semiclassical description, there are non-negligible correlations between the state of the ion and that of the cavity also well within the regime where the ion is located at the trap center. 
One could then conclude that the semiclassical approximation as in \cite{cormick2012structural} is not appropriate for any of the parameter ranges in our studies. Nevertheless, the semiclassical treatment predicts correctly the parameter range for the optomechanical transition.

Summing up, the study of the transition in the location of an ion induced by the dispersive coupling to a cavity field shows, in the regime with around ten photons, some features of the behaviour predicted by the semiclassical treatment in \cite{cormick2012structural}, but does not display indications of metastability. We confirm the expectation that the transition becomes sharper when the overlap between the two competing equilibrium configurations decreases. We also observe local maxima of logarithmic negativity, mutual information and von Neumann entropy of the cavity in the transition region. We do not find, however, a decrease in the spectral gap, so that at all times the asymptotic state is clearly separated from the rest. 

\section*{Acknowledgments}

A. K. acknowledges funding from a CONICET fellowship.

\appendix

\section{Semiclassical analysis of the equilibrium configurations} \label{sec:classical_stability}

In this section we provide more details regarding the classical equilibrium configurations. Using (\ref{xeq}) and defining $z=1-(\overline{x}/x_{\rm eq})^2$, the equation that determines the equilibrium positions at the sides is:
\begin{equation}\label{eq_cuartico}
  1 + C^2 \left(z^2- 1 + c \right)^2  = 4\, C \gamma \, z
\end{equation}
where we use the dimensionless parameter combinations:
\begin{equation}
 c=-\frac{\Delta_c}{U_0} +1
\end{equation}
associated with the detuning, and
\begin{equation}
\gamma= \left(\frac{x_\omega}{x_{\rm eq}}\right)^2 \frac{\eta^2}{\omega \kappa} 
\end{equation}
associated with the pumping strength. These solutions exist only for strong enough pumping, and when they exist, they are stable.

By looking at the second derivative of the total effective potential $V=V_{\rm eff}+ V_{\rm ion}$ one can determine conditions for the
system parameters for which the semiclassical solutions $\overline x$ transition from stable to unstable equilibrium configurations.
For the solution at the origin the critical condition is given by the vanishing of the second derivative, corresponding to:
\begin{equation}\label{segunda_estrellita}
  \gamma_c^{(0)} = \frac{1+ c^2 C^2}{4C} 
\end{equation}

For the equilibrium solutions at the sides, the critical parameter combination can be found from putting together eq. (\ref{eq_cuartico}) with 
\begin{equation} \label{eq:side_critical}
z\left(z^2+c-1 \right) = \frac{\gamma}{C} .
\end{equation}
We could not derive an explicit general formula for the critical value of $\gamma$ corresponding to the appearance of the equilibrium positions at the sides. However, for the particular case of zero detuning, $c=1$, the equations greatly simplify and one can find:
\begin{equation}\label{primera_estrellita}
  \gamma_c^{(s)} = \frac{1}{3^{3/4} \sqrt{C}}
\end{equation}
where we use the superindex $(s)$ to indicate that this critical value corresponds to the equilibrium at the sides.

\section{Gaussian states} \label{sec:gaussian}

Let us consider a set of dimensionless canonical operators $ q_k,p_k $ representing a system of $N$ degrees of freedom.
The operators $ q_k,p_k $ are related to creation and annihilation operators of a given mode
through \cite{ekert1995}:

\begin{equation}\label{canonicos}
\begin{aligned}
  q_k & = \frac{(a_k + a_k^\dagger)}{\sqrt{2}}\\
  p_k & = \frac{i(a_k^\dagger - a_k)}{\sqrt{2}}
\end{aligned}
\end{equation}

If canonical operators are grouped in a vector as
\begin{equation}\label{primeros.momentos}
  \vect{R} = \left( q_{1},p_{1},q_{2},p_{2},...,q_{N},p_{N}  \right),
\end{equation}

\noindent the canonical commutation relations can be recast as:
\begin{equation}\label{conmutacion}
  \conm{R_k}{R_l} = i \Omega_{kl},
\end{equation}
\noindent where $\Omega$ is the symplectic form:

\begin{equation}
  \Omega = \bigoplus_{k=1}^{N} \omega, \qquad \omega= \begin{pmatrix}
                                                     0 & 1 \\
                                                    -1 & 0 \\
                                                  \end{pmatrix}
\end{equation}

The set of displacement operators $D(\boldsymbol{\alpha})$ with $\boldsymbol\alpha \in \mathbb{C}^N$ is complete,
in the sense that any operator $O$ acting on the Hilbert space $\mathcal{H}$ can be written as \cite{cahill1969}:

\begin{equation}\label{func.caract}
  O = \int_{\mathbb{C}^N} \frac{d^{2N} \boldsymbol \alpha}{\pi^N} \text{Tr}\left[O D(\boldsymbol\alpha) \right]
       D^\dagger (\boldsymbol\alpha)
\end{equation}

Function $\chi_o(\boldsymbol\alpha) = \text{Tr}\left[O D(\boldsymbol\alpha) \right]$ is termed
the characteristic function of the operator $O$. 
The set of Gaussian states is the set of states with Gaussian characteristic functions, and thus
Gaussian states are completely characterized by their first and second moments, that is, by 
the vector of first moments $P$ (\ref{pm}) and by the covariance matrix $\sigma$ (\ref{cm}):

\begin{equation}\label{pm}
  P_i = \langle R_i \rangle
\end{equation}
\begin{equation}\label{cm}
  \sigma_{ij} = \frac{1}{2} \langle \lbrace R_i, R_j \rbrace \rangle  - \langle R_i \rangle \langle R_j \rangle
\end{equation}
This permits the description of continous variable systems in terms of finite matrices. Particular examples of Gaussian states 
are coherent states, thermal states of quadratic Hamiltonians, and squeezed states \cite{ferraro2005}.

There are several quantities to measure the Gaussian character of an arbitrary state, based on the relative entropy \cite{genoni2008, marian2013}, Hilbert-Schmidt norm \cite{genoni2007} or Bures metric \cite{marian2013_ghiu}. 
We consider a non-Gaussianity measure based on the Wehrl entropy, proposed and studied in \cite{solomon2012}.
Given a state $\rho$, the non-Gaussianity measure $\mathcal N (\rho)$ is defined as the difference of Wehrl entropies:

\begin{equation}\label{non_gaussianity_wehrl}
  \mathcal N (\rho) = H_W(\rho_G) - H_W(\rho)\,.
\end{equation}

\noindent Here, the reference state $\rho_G$ 
represents the Gaussian state with the same first and second moments as $\rho$, and the Wehrl entropy for an $N$-mode state is defined as \cite{solomon2012}

\begin{equation}\label{wehrl_entropy}
  H_W(\rho) = - \frac{1}{\pi^N} \int \prod_{j=1}^N d^2 \alpha_j Q_\rho (\alpha) \log  Q_\rho (\alpha) \,.
\end{equation}

\noindent This entropy is based on the Husimi quasi-probability distribution $Q_\rho(\alpha)=\bra{\alpha} \rho \ket{\alpha}$, with $\alpha \in \mathbb C^N$, which is defined analogously for the subsystem states. We note that we follow the definition of $Q$ in \cite{solomon2012}, which is normalized to $\pi$ instead of 1 \cite{LEE}.

The measure $\mathcal N (\rho)$ is always non-negative, being 0 if and only if the state $\rho$ is Gaussian. 
This measure also has the desirable property that if two quantum states possess phase-space representations as given by the Husimi functions that are related by a uniform scaling of all phase space coordinates, then they are assigned the same amount of non-Gaussianity \cite{solomon2012}.

\section{Numerical methods to find the steady state}
\label{sec:methods}

The Lindblad master equation is of the form $\dot \rho = \mathcal{L}(\rho)$ and includes the Hamiltonian terms and one dissipative channel associated with cavity losses.
Since $\mathcal{L}$ is time-independent, at least one steady state exists \cite{rivas2012}.
If $\mathcal{L}$ is diagonalizable, its eigenvectors, defined by $\mathcal{L}\rho_i = \nu_i \rho_i$, satisfy $\text{Re} \left[ \nu_i \right] \le 0 $.
The set of eigenvectors with eigenvalues satisfying  $\text{Re}\left[ \nu_i \right]=0$ form the decoherence free subspace of $\mathcal{L}$ \cite{halati2019}.
If the zero eigenvalue is degenerate with degeneracy $n$, 
then there is an inifinite set of states towards which the system can relax depending on the initial state. However, in absence of symmetries the steady state is generally unique  \cite{ciuti2018, jiang2014} and given by the right eigenvector corresponding to the 0 eigenvalue: $ \rho_{ss} = \rho_0 / \text{Tr}\left[ \rho_0 \right]$.  The Liouvillian gap $\abs{\text{Re}\left[ \nu_1 \right] }$, also termed asymptotic decay rate, determines the system's slowest relaxation rate \cite{ciuti2018}.

For our numerical calculations, we describe the state of the system in a truncated basis of Fock states for each subsystem, i.e. eigenstates of the photon number operator for the cavity and the vibrational number operator corresponding to the trap potential alone for the motion. We note that this choice becomes rather inefficient in the limit with the ion well localized at the sides; however, our main focus is the bistable region and we need to describe the three equilibrium locations of the ion at the same time. In contrast, the number of states required to describe the cavity can be reduced applying a displacement operator on the basis of Fock states. This is particularly useful for high mean photon numbers and will be explained in more detail below.

%The cavity basis cutoff varies with every different value of $\kappa$, and is chosen in such a way that the density matrix populations of each subsystem decay below a threshold that we chose as $10^{-5}$ in the worst case. However, for most of the different parameter sets the truncated populations are several orders of magnitude below this threshold, and we checked that the results do not vary significantly if this threshold is made smaller. 

We write the state in vectorized form, so that the action of the evolution superoperator $\mathcal{L}$ corresponds to multiplication by a matrix $L$ \cite{ciuti2018,jiang2014}. Since the cutoff dimensions needed to correctly represent the states are too big in the composite Hilbert space, it is not feasible to solve the diagonalization of the whole Liouvillian operator by means of an exact diagonalization algorithm. Hence we resort to the Arnoldi algorithm for finding the eigenvalues in the spectral region of interest \cite{lohoucq1996}.

The region of interest in the spectrum of $L$ corresponds to the eigenvalues with smallest real part, and in particular also the null eigenvalue, while the Arnoldi method is efficient for finding eigenvalues of largest magnitude in the spectrum.
Because of this, we solve the equivalent problem of finding the eigenvectors of the dynamical map $\Lambda(t)=\exp(L t)$ for some fixed short time $t$. 
In this case, multiplication of the map by a vector $x_0$ amounts to evolution of the initial point $x_0$ as given by $\dot x = L x$.
Then the region of interest in the spectrum consists of the eigenvalues of the map with largest absolute value, and particularly the eigenvalue 1, corresponding to the steady state.

Since the map eigenvectors are time independent, the Lindblad equation to obtain the map $\Lambda(t)$ can be integrated by means of the Euler method at very short times $\Delta t$, so that $t = N_{\rm steps} \Delta t$. For our calculations, we took a step $\Delta t = 10^{-4} /\omega$, which provided sufficient precision without requiring exceedingly long computation times. 

This short integration time step is essential to achieve a high precision for the eigenvectors. Since this is prohibitive in the original basis of Fock states of the cavity we perform a change to a basis consisting of displaced Fock states. In order to do this, we first obtain preliminary results in the original basis at low precision, where we verify that the mean values already converge by varying the basis cutoff. We choose the maximum basis cutoff of these preliminary results such that the reduced density matrix populations decay below a threshold value of $10^{-5}$ in the worst case. From this solution, we find an estimation for the mean value of the cavity quadratures. Then we perform the change of basis to a displaced number state basis for the cavity field system, which greatly reduces the number of states in our description. We keep the cutoff in this new basis fixed, and augment the precision by decreasing the integration time. Here we verify that the mean values are still fixed to the same values obtained previously but the precision of the eigenvectors thus obtained is now increased as to assure the validity of the remaining numerical calculations.

%%---- Para bibtex-----
%\bibliographystyle{ieeetr}       % Set the bibliography style to AMS
%\bibliography{biblio}{}

%%---- Para biblatex-----
% \printbibliography

\end{document}